\documentclass[acus]{JAC2003}
\usepackage{graphicx}
\begin{document}
\title{
\textmd{\normalsize PAC 2003 Particle Accelerator Conference, 
Portland, Oregon (May 12--16) TPPB092} \\
A VERY FAST RAMPING MUON SYNCHROTRON\\ FOR A  NEUTRINO FACTORY}
\author{D. J. Summers\thanks{summers@relativity.phy.olemiss.edu}, 
University of Mississippi--Oxford, University, MS 38677, USA\\ 
J. S. Berg, R. B. Palmer, Brookhaven National Laboratory, Upton, NY 
11973, USA\\ 
A. A. Garren, University of California, Los Angeles, CA 90095, USA}
\maketitle

\begin{abstract}
A 4600 Hz fast ramping synchrotron is studied as
an economical way of accelerating muons from 4 to
20 GeV/c for a neutrino factory. Eddy current
losses are minimized by the low machine duty
cycle plus thin grain oriented silicon steel
laminations and thin copper wires. Combined
function magnets with high gradients alternating
within single magnets form the lattice.
Muon survival is 83\%.
\end{abstract}

\bigskip

Historically synchrotrons have provided economical particle
acceleration. Here we explore a very fast ramping muon synchrotron 
\cite{nufact02} for a
neutrino factory \cite{factory}. The accelerated muons are stored in a
racetrack to produce neutrino beams
($\mu^- \to e^- \, {\overline{\nu}}_e \, \nu_{\mu}$ \, or \, 
$\mu^+ \to e^+ \, \nu_e \, {\overline{\nu}}_{\mu}$). Neutrino oscillations
\cite{oscillation} have been observed at experiments \cite{homestake} 
such as Homestake, Super--Kamiokande, SNO, and KamLAND. Further
exploration using a neutrino factory could reveal CP 
violation in the lepton sector.

This synchrotron must accelerate muons from 4 to 20 GeV/c
with moderate decay loss ($\tau_{\mu^{\pm}}$ = 2.2 $\mu$S). 
Because synchrotron radiation goes as $m^4$,
muons radiate two billion times (\,$(105.7/.511)^4$\,) 
less power than electrons for a given ring
diameter and lepton energy. 
Grain oriented silicon steel (3\% Si) is used to provide a high
magnetic field with a high $\mu$ to minimize magnetic energy ($B^2/2\mu$)
stored in the
yoke. Magnetic energy stored in the gap is minimized by
reducing its size. Cool muons \cite{cool} with low beam emittance  
allow this. 
The voltage needed to drive a magnet is 
$-L \, di/dt$. 
Voltage is minimized by shrinking the volume of 
stored energy to reduce the
inductance, $L$. 

Acceleration to 4 GeV/c might feature fixed field dogbone arcs \cite{dogbone} to
minimize muon decay loss. Fast ramping synchrotrons \cite{dogbone, snowmass}
might also accelerate cold muons to higher energies 
for a $\mu^+ \, \mu^-$ collider \cite{collider}.


We form arcs with sequences of combined function cells
within continuous long magnets, whose poles are alternately shaped to give
focusing gradients of each sign. A cell has been simulated
using SYNCH \cite{synch}. Gradients alternate from
positive 20 T/m gradient (2.24 m long), to zero gradient (.4 m long) to
negative 20 T/m gradient (2.24 m) to zero gradient (0.4 m), etc. 
Details are given in Table 1.

\begin{table}[!t]
\begin{center}
\caption{Combined function magnet cell parameters. Five cells make up an
arc and 18 arcs form the ring.}
\vspace*{2mm}
\begin{tabular}{lcc}
\hline
Cell length &m&5.28\\
Combined Dipole length &m& 2.24\\
Combined Dipole B$_{\rm central}$ &Tesla& 0.9\\
Combined Dipole Gradient &T/m& 20.2\\
Pure Dipole Length &m& 0.4\\
Pure Dipole B &Tesla& 1.8\\
Momentum &GeV/c&20\\
\hline
Phase advance/cell&deg&72\\
beta max &m& 8.1\\
Dispersion max &m&0.392\\
\hline
Normalized Trans.~Acceptance & $\pi$ mm rad & 4 \\
\hline
\end{tabular}
\end{center}
\vspace*{-8mm}
\end{table}

\begin{figure}[!b]
\vspace*{-1mm}
\centering
\includegraphics*[width=83mm]{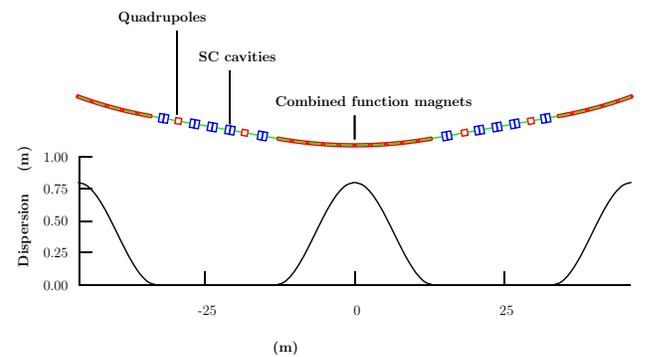}
\caption{\label{label1}
Combined function magnets bend the muons in the arcs.  Superconducting
RF cavities accelerate muons in the straight sections. Two quadrupoles per
straight section provide focusing. The straight sections are dispersion free.}
\end{figure}

It is proposed to use 5 such arc cells to 
form an arc segment. These segments are alternated with straight 
sections containing RF. The phase advance through one arc segment is 5 x 
72$^0$ = 360$^0$. This being so, dispersion suppression between 
straights and arcs can be omitted. With no dispersion in the straight 
sections, the dispersion performs one full oscillation in each arc 
segment, returning to zero for the next straight as shown in Fig.~1. 
There will be 18 such arc segments and 18 straight sections, forming 
18 superperiods in the ring.
Straight sections (22 m) without dispersion are used for superconducting 
RF, and, in two longer straights (44 m), the injection and extraction. 
To assure sufficiently low magnetic fields at the cavities, 
relatively long field free regions are 
desirable. A straight consisting of two half cells 
would allow a central gap of 10 m between quadrupoles, and two 
smaller gaps at the ends. 
Details are given in Table 2.
Matching between the arcs and straights is not yet designed.
The total circumference of the ring including combined functions magnets
and straight sections adds up to 917 m 
($18 \times 26.5 \, + \, 16 \times 22 \, + \, 2 \times 44$).


The RF (see Table 3) must be distributed around the ring to 
avoid large differences between the beam momentum (which increases in 
steps at each RF section) and the arc magnetic field (which 
increases continuously). 
The amount of RF used is a tradeoff between cost and muon survival. Survival
is somewhat insensitive to the fraction of stored energy the beam
removes from the RF cavities, because the voltage drop is offset by time
dilation.

\begin{table}[!t]
\begin{center}
\caption{Straight section lattice parameters.}
\vspace*{2mm}
\tabcolsep=2mm
\begin{tabular}{ccccc}
\hline
$\phi$ & L$_{\rm cell}$/2 & L$_{\rm quad}$ & dB/dx  & a \\
77$^{\,0}$  & 11 m & 1 m & 7.54 T/m & 5.8 cm \\ 
\vspace*{-6pt} & & & &  \\
$\beta_{\rm max}$ &$\sigma_{\rm max}$ & B$_{\rm pole}$ & U$_{\rm mag}$/quad & \\
36.6 m &.0195 m & 0.44 T & $\approx$ 3000 J & \\ \hline
\end{tabular}
\end{center}
\end{table}

\begin{table}[!b]
\begin{center}
\caption{Superconducting RF parameters.}
\vspace*{2mm}
\tabcolsep=2mm
\begin{tabular}{lcc}
\hline
Frequency                 & 201                & MHz      \\
Gap                       & .75                & m       \\
Gradient                  & 15                 & MV/m     \\
Stored Energy             & 900                & Joules   \\
Muons per train           & $5 \times 10^{12}$ &          \\
Orbits (4 to 20 GeV/c)    & 12                 &          \\
No.~of RF Cavities        & 160                &          \\
RF Total                  & 1800               & MV       \\
$\Delta$U$_{\hbox{beam}}$ & 110                & Joules   \\
Energy Loading            & .082               &          \\
Voltage Drop              & .041               &          \\
Muon Acceleration Time    & 37                 & $\mu$sec \\
Muon Survival             & .83                &          \\ \hline            
\end{tabular}
\end{center}
\end{table}


The muons accelerate from 4 to 20 GeV.  If they are extracted at
95\% of full field they will be injected at 19\% of full field.
For acceleration with a plain sine wave, injection occurs at
11$^0$ and extraction occurs at 72$^0$.  So the phase must
change by 61$^0$ in 37 $\mu$sec.  Thus the sine wave goes through
360$^0$ in 218 $\mu$sec, giving 4600 Hz.
 
  Estimate the energy stored in each 26.5 m long combined function magnet.
The gap is about .14 m wide and has an average height of .06 m. Assume an
average field of 1.1 Tesla. The permeability constant, $\mu_0$, is $4\pi\times
10^{-7}$. $W = {B^2 / {2{\mu_0}}}[\hbox{Volume}] =$ 110\,000 Joules. Next
given one turn, an LC circuit capacitor, and a 4600 Hz frequency; estimate
current, voltage, inductance, and capacitance.

\begin{eqnarray}
B = {{\mu_0\,NI}\over{h}}  \, \rightarrow\,
I = {{Bh}\over{\mu_0\,N}} = 52 \, \hbox{kA} \\
W = .5\,L\,I^2  \, \rightarrow\,  L = {2\,W / {I^2}} =
80\,\mu\hbox{H} \\
f = {1\over{2\pi}}\sqrt{1\over{LC}}  \, \rightarrow\, 
C = {1\over{L\,(2\pi f)^2}} = 15\, \mu\hbox{F} \\ 
W = .5\,C\,V^2  \, \rightarrow\,  V = \sqrt{2W / {C}} = 120\,\hbox{kV}
\end{eqnarray}
 
Separate coils might be put around each return yoke to halve the voltage.
The stack of SCRs driving each coil might be center
tapped to halve the voltage again. Four equally spaced 6 cm 
coil slots could be created
in each side yoke using 6 cm of wider laminations 
to cut the voltage by five, while leaving the pole faces
continuous. 6 kV is easier to insulate than 120 kV. It may be useful to shield
\cite{nufact02} or chamfer \cite{school} magnet ends to avoid large eddy 
currents where the
field lines typically do not follow laminations. A DC offset power supply
could be useful \cite{nufact02}. Neutrino horn power supplies 
look promising.

Grain oriented silicon steel is chosen for the return yoke due to its
high permeability at high field at noted in Table 4.  
The skin depth \cite{lorrain}, $\delta$,  
of a lamination is 160 $\mu$m from eqn.~5. $\rho$ = resistivity.

\begin{equation} 
{\delta = \sqrt{\rho \, / \, \pi \, f \,\mu} = 
\sqrt{47\times{10^{-8}} \, / \, \pi \, 4600 \, 1000 \, \mu_0}}
\end{equation}

Take $\mu = 1000 \mu_0$ as a limit on magnetic saturation and hence energy
storage in the yoke. Next estimate the fraction of the inductance of the yoke
that remains after eddy currents shield the laminations \cite{lucent}.
The lamination thickness, $t$, equals 100 $\mu$m \cite{arnold}.  
L/L$_0$ =
$(\delta/t) \, (\sinh(t/\delta) + \sin(t/\delta)) \,  / \,
(\cosh(t/\delta) + \cos(t/\delta))$ = 0.995.
So it appears that magnetic fields can penetrate 100 $\mu$m thick laminations 
at 4600 Hz.  Thicker 175 $\mu$m thick laminations \cite{armco} would be 
half as costly and can achieve
a somewhat higher packing fraction. 
L/L$_0$($t$ = 175 $\mu$m) = 0.956. 

\begin{table}[!b]
\begin{center}
\caption{Permeability ($B/\mu_0H$).     
Grain oriented silicon steel has a far
higher permeability parallel ($\parallel$) to 
than perpendicular ($\perp$) to
its rolling direction \cite{armco,ferro}. T = Tesla.}
\vspace*{2mm}
\renewcommand{\arraystretch}{1.05}
\tabcolsep=3mm
\begin{tabular}{lrrr} \hline 
Material                    &  1.0 T & 1.5 T & 1.8 T \\ \hline 
1008 Steel                  &    3000 &   2000 &  200   \\
Grain Oriented ($\parallel$)&   40000 &  30000 & 3000   \\
Grain Oriented ($\perp$)    &    4000 & 1000   &        \\
\hline 
\end{tabular}
\end{center}  
\end{table}

Calculate the resistive energy loss in the copper coils.
There are four 5\,cm square copper conductors each
5300\,cm long.
R = ${5300 \ (1.8\,\mu\Omega\hbox{-cm}) \, / \, {(4) \, (5^2)}} = 
95\,\mu\Omega.$ So, 
$P = I^2R\int_0^{2\pi}\!\cos^2(\theta)\,d\theta = \hbox{130\,000 w/magnet}.$
Eighteen magnets give a total loss of 2340 kW.
But the neutrino factory runs at 30 Hz.  Thirty half cycles 
of 109 $\mu$sec per second gives a duty factor of 300 and a total $I^2R$ loss
of 8000 watts.  Muons are orbited in opposite directions on alternate cycles. 
If this proves too cumbersome, the duty cycle factor could be lowered to 150.
See if .25 mm (30 gauge) wire is usable.
From eqn.~6, the skin depth, $\delta$, of copper at 4600 Hz is 0.97 mm. 

\begin{equation}
\delta = \sqrt{\rho \, / \, \pi \, f \,\mu_0} = 
\sqrt{1.8\times{10^{-8}} \, / \, \pi \, 4600 \, \mu_0}
\end{equation}
 
Now calculate the dissipation due to eddy currents in this $w$ = .25 mm wide
conductor, which will
consist of transposed strands to reduce this loss \cite{school, sasaki}.
To get an idea, take the maximum B-field
during a cycle to be that generated by a 0.025m radius conductor carrying
26000 amps.
The eddy current loss in a rectangular conductor made of transposed square
wires .25 mm wide (Litz wire \cite{mws}) 
with a perpendicular magnetic
field is as follows. 
$B = {{\mu_0\,I}/{2\pi r}} = 0.2$ Tesla.

\begin{equation}
P = \hbox{[Volume]}{{(2\pi\,f\,B\,w)^2}\over{24\rho}}
\end{equation}

\begin{equation}
P = [4 \ .05^2 \ 53]\, {{(2\pi \ 4600 \ .2 \ .00025)^2} \over
{(24)\,1.8\times{10^{-8}}}} = 2800 \
\hbox {kW} 
\end{equation}

Multiply by 18 magnets and divide by a duty factor of 300
to get an eddy current loss in the copper of 170 kW.
Stainless steel water cooling tubes will dissipate a similar amount
of power \cite{dogbone}. Alloy titanium cooling tubes would dissipate half
as much.

Do the eddy current losses \cite{sasaki} in the 100 $\mu$m thick iron 
laminations. Use eqn.~7 with 
a quarter meter square area, a 26.5 m length, and
an average field of 1.1 Tesla.

\begin{equation}
\hbox{P}
=  [(26.5) \, \, (.5^2)]\,
{{(2\pi \ 2600 \ 1.1 \ .0001)^2} \over
{(24)\,47\times{10^{-8}}}}
=  5900 \
\hbox {kW}
\end{equation}

Multiply by 18 magnets and divide by a duty factor of 300 to get an
eddy current loss in the iron laminations of 350 kW or 700 watts/m
of magnet. 
So the iron will need some cooling. The ring only ramps 30 time per second, so
the $\int{\bf{H}}{\cdot}d\,{\bf{B}}$ hysteresis losses will be low, even
more so because of the low coercive force (H$_c$ = 0.1 Oersteds) of grain 
oriented silicon
steel. This value of H$_c$ is eight times less than 1008 low carbon steel.


The low duty cycle of the neutrino factory leads to eddy
current losses of less than a megawatt in a 4600 Hz, 917 m circumference ring. 
Muon survival is 83\%. The high
permeability of grain oriented silicon steel permits high fields with little
energy stored in the yoke. Gradients are switched within dipoles to minimize
eddy current losses in ends. Time dilation allows extra orbits with little
muon decay at the end of a cooling cycle. This 
allows one to use more of the stored RF energy. 
Much of the magnetic field in our lattice is used for focusing rather than
bending the muon beam. More muon cooling would lead to less
focusing, more bending, and an even smaller ring.

This work was supported by the U.~S.~DOE and NSF. 
Many thanks to K.~Bourkland, S.~Bracker \cite{lasker}, C.~Jensen,
S.~Kahn, H.~Pfeffer, G.~Rees, Y.~Zhao, and M.~Zisman.


\end{document}